\documentclass{iopart}
\usepackage{iopams}
\usepackage{amsmath}
\usepackage{amsfonts}
\usepackage{latexsym}
\usepackage{euscript}
\usepackage[dvips]{graphicx}

\begin{document}
\title{Transfer of linear momentum from the quantum vacuum to a magnetochiral molecule}
\author{M. Donaire$^{1}$, B.A. van Tiggelen$^{2}$ and G.L.J.A. Rikken$^{3}$}
\address{$^{1}$Laboratoire Kastler-Brossel, CNRS, ENS  and UPMC, Case 74, F-75252 Paris, France}
\address{$^{2}$Universit\'{e} Grenoble 1/CNRS, LPMMC UMR 5493, B.P.166, 38042 Grenoble, France}
\address{$^{3}$LNCMI, UPR 3228 CNRS/INSA/UJF Grenoble 1/UPS, Toulouse \& Grenoble, France}
\ead{donaire@lkb.upmc.fr}

\begin{abstract}\\
In a recent publication \cite{PRLDonaire} we have shown using a QED approach that, in the
presence of a magnetic field, the quantum vacuum coupled to a chiral
molecule provides a kinetic  momentum directed along
the magnetic field. Here we explain the physical mechanisms which operate in the
transfer of momentum from the vacuum to the
molecule. We show that the variation of the molecular kinetic energy originates from the magnetic energy associated with the vacuum
correction to the magnetization of the molecule. We carry out a semiclassical calculation of the vacuum momentum
and compare the result with the QED calculation.
\end{abstract}
\submitto{Journal of Physics: Condensed Matter}
\maketitle

\section{Introduction}
It is well known that the quantum fluctuations of the electromagnetic (EM)
field coupled to electric charges generate an observable interaction
energy \cite{Casimir,Lamoreaux,Milonnibook}. The fluctuations which mediate the self-interaction of
electrons bound to atomic nuclei give rise to the Lamb shift of atomic levels; the
fluctuations which mediate the interaction between nearby molecules
generate van-der-Waals energies; and finally the fluctuations between
macroscopic dielectrics generate the Casimir energy. Direct
observation of these energies is possible by spectroscopy, atomic interferometry or nanomechanical means \cite{Alex,Gorza,Bressi,Capasso}.\\
\indent Less well-known is the fact that other observable quantities, functions
of the EM field, can be influenced by quantum fluctuations under
certain symmetry conditions. That is, when the time-space symmetries
of the medium to which the fluctuations couple are compatible with
the symmetries of some observable operator, the expectation value of
that operator in the vacuum state of the system medium-EM field may take a non-zero value.
This is the case of the linear momentum of the EM field when quantum
fluctuations couple to a medium in which both parity (P) and
time-reversal (T) symmetries are broken. In particular, the constitutive equations of a magneto-electric medium contain a
non-reciprocal (\emph{nr}) EM susceptibility \cite{Barron84}, $\chi_{EM}^{nr}$, which results from broken P and T. Generally $\chi_{EM}^{nr}$ is an
antisymmetric T,P-odd tensor which generates an electric polarization in response
to a magnetic field, $\Delta\mathbf{P}=\chi^{nr}_{EM}\cdot\mathbf{B}$, and conversely, a magnetization as a response to an electric field,
$\Delta\mathbf{M}=-\chi^{nr}_{EM}\cdot\mathbf{E}$.
As a result of the matter-field coupling
and  momentum conservation, linear momentum
can be transferred from the EM vacuum to matter during the process that controls the
break down of the symmetries. A nonzero tensor $\mathbb{\chi}_{EM}^{nr}$ is found in any medium in crossed external electric and magnetic fields, in a moving dielectric medium and in a chiral medium exposed to a magnetic field \cite{vTg2008}. In this article we concentrate on the latter case.\\
\indent It is a generic phenomenon in field theory that the breakdown of a symmetry is accompanied by a non-zero vacuum expectation value (VEV)
of some physical observable associated to the symmetry. In our case the P and T symmetries happen to be broken explicitly by the presence of a chiral 
molecule and the action of an external magnetic field, $\mathbf{B}_{0}$. Correspondingly, a non-zero  VEV of the EM momentum shows up in the direction
along which the symmetries are broken, $\mathbf{B}_{0}$. The question arises whether it could be possible to take advantage of this phenomenon for 
practical purposes.  To this end we will show that, due to the conservation of total linear momentum, there exists necessarily a transfer of kinetic momentum to the chiral molecule of equal magnitude and opposite sign to the VEV of the EM momentum.\\ 
\indent Also, in the context of high energy physics,  it is known that the Electro-Weak interaction violates the P and T symmetries \cite{Wu,Alavi}, 
and CP (i.e., T) is also expected to be naturally broken in QCD. In the latter case, the existence of a light pseudo-scalar particle, the axion, has been postulated as a solution 
of the so-called \emph{strong CP problem} \cite{Perccei-Quinn}, and indirect observations of the axion through its coupling to the EM field have been 
suggested \cite{Zavattini} and put into practice \cite{PVLAS}. In the PVLAS experiment  \cite{PVLAS} signatures of the axion-photon interaction are
investigated through the anomalous rotation of the polarization direction of optical light in the presence of an intense magnetic field, similar to 
the vacuum birefringence effect expected from the polarization of the QED vacuum \cite{Klein,Rizzo}. 
Nonetheless the rotation of the polarization axis of light may be caused by the break down of the P or T symmetries separately, while the effect
described in the present article needs  the simultaneous breakdown of both symmetries and is independent of polarization.\\
\indent Already in classical electrodynamics, a kinetic momentum is acquired by matter under
external time-dependent electric and magnetic fields perpendicular to each other \cite{Rikken1,Rikken2}. Conservation of total momentum implies that the
classical EM momentum and the kinetic momentum
must be equal in magnitude and opposite in sign. Several expressions can be found in the
literature for this classical EM momentum, all derived from some kind of phenomenological assumptions. This controversy is known
as the Abraham-Minkowski problem. According to Abraham's prescription, the density of EM momentum is $\mathbf{G}_{A}=c^{-2}\mathbf{E}\wedge\mathbf{H}$, while
Minkowski's formula reads $\mathbf{G}_{M}=\mathbf{D}\wedge\mathbf{B}$. More recently, the microscopical approach of Nelson \cite{Nelson}, which appeals to
the Lorentz force over bound and free charges and currents for the coupling between matter and radiation, predicts a momentum density of an EM wave in a material
medium, $\mathbf{G}_{N}=\epsilon_{0}\mathbf{E}\wedge\mathbf{B}$.\\
\indent Starting with a classical Lagrangian
that reproduces Maxwell's equations and using the constitutive equations of a homogeneous and isotropic medium of density $\rho$ moving at velocity $\mathbf{v}$, Feigel \cite{Feigel}
found a total conserved momentum density $\mathbf{G}_{M}=\rho\mathbf{v}+\mathbf{G}_{A}$, from which the kinetic momentum density of matter is found,
$\rho\mathbf{v}=\epsilon_{0}\frac{\epsilon_{r}\mu_{r}-1}{\mu_{r}c}\mathbf{E}\wedge\mathbf{B}$ \cite{Rikken1}.
However, following an approach close to Nelson's, van Tiggelen  has arrived to another conservation law \cite{vTg2008}. On the one hand, the combination of
Maxwell's equations and the constitutive equations for a moving medium of mass density $\rho_{m}$ yields, $\partial_{t}(\mathbf{D}\wedge\mathbf{B})=\nabla\cdot\mathbb{T}-\mathbf{F}_{L}$,
 with $\mathbb{T}_{ij}=(B_{i}H_{j}+D_{i}E_{j})-(\epsilon E^{2}+\mu^{-1}B^{2}+\rho_{m}v^{2})\delta_{ij}/2$ and
 $\mathbf{F}_{L}=\rho_{q}\mathbf{E}+\mathbf{J}\wedge\mathbf{B}$ the Lorentz force density, $\rho_{q}$ being the density of charge. On the other hand,
 Newton's law reads $\partial_{t}(\rho\mathbf{v})+(\mathbf{v}\cdot\nabla\rho_{m})\mathbf{v}=\mathbf{F}_{L}$. Lastly,
 combining both equations and integrating in space one obtains the total conserved momentum
 $m\mathbf{v}+\int\textrm{d}^{3}r\mathbf{G}_{N}$. The equation of motion which derives from this approach is \cite{Rikken1,Nelson,Kawka2}
 $\rho_{m}\mathbf{v}=\epsilon_{0}(\epsilon_{r}-1)\mathbf{E}\wedge\mathbf{B}$.\\
\indent In addition to the classical EM momentum coming from external crossed fields, EM quantum fluctuations can generate extra terms
when the constitutive equations contain non-reciprocal susceptibilities, $\chi_{EM}^{nr}$ \cite{vTg2008,Feigel,Croze}. The basic argument is that
a non-zero $\chi_{EM}^{nr}$ generates what is called spectral non-reciprocity. That is, there exists a difference in the frequency of normal modes propagating in
opposite directions along the axis where the P symmetry is broken, $\tilde{\omega}(\mathbf{k})\neq \tilde{\omega}(-\mathbf{k})$. This implies that the addition of
the linear momenta propagated by these modes do not cancel out.
The total momentum of the normal modes is the momentum of the vacuum fluctuations and therefore it is referred to as Casimir momentum.\\
\indent As pointed out in Refs.\cite{vTg2008,KrsicvTgRikken} Feigel's approach suffers from a number of problems. In the first place,
since $\mathbf{E}\wedge\mathbf{H}$ is proportional to the  energy current, one expects that in thermodynamic equilibrium $\langle\mathbf{G}_{A}\rangle=\mathbf{0}$, and so
no variation of the kinetic momentum can be obtained from variations of $\langle\mathbf{G}_{A}\rangle$. Second, the final result obtained by Feigel for the Casimir momentum
density, $\sim\hbar\int\chi_{EM}^{nr}k^{3}\textrm{d}k$, seems to
lack a reference frame since this equation is not Lorentz or even Galilean invariant \cite{KrsicvTgRikken}.\\
\indent In addition, the derivations of the above
expressions for the conserved linear momentum are purely phenomenological and lack an explicit quantum
interaction Hamiltonian between matter and radiation. Also, a macroscopic approach can easily lead to an erroneous prediction. For instance, it is shown
in Ref.\cite{vTg2008} that for
a chiral distribution of electrically polarizable molecules in an external magnetic field,
$\langle\mathbf{E}(\mathbf{r})\wedge\mathbf{B}^{*}(\mathbf{r})\rangle=\mathbf{0}$ should hold
locally \footnote{The reason for this result is basically that $\mathbf{B}(\mathbf{r})=\mu_{0}\mathbf{H}(\mathbf{r})$ for a non-magnetic medium, and
the energy flow $\langle\mathbf{E}(\mathbf{r})\wedge\mathbf{H}^{*}(\mathbf{r})\rangle$ vanishes.}, although the medium contains a nonzero nonreciprocal effective
(\emph{eff}) response which would make
$\langle\mathbf{E}_{eff}\wedge\mathbf{B}_{eff}^{*}\rangle\neq\mathbf{0}$. Another problem that should be resolved in a quantum microscopic treatment is the UV divergence
of the expression for the quantum EM momentum density obtained in a homogeneous medium
\cite{Feigel}. 
This line of investigation was started by Kawka, van-Tiggelen and Rikken
in Refs.\cite{Kawka2,Kawka1}, where they computed the Casimir momentum of an atom in crossed external electric and magnetic fields. The divergences were found there to disappear by mass renormalization and the leading quantum contribution was found to be a factor $\alpha^{2}$ smaller than the classical EM momentum.\\
\indent In Ref.\cite{PRLDonaire} we reported the quantum computation of  the Casimir momentum for a chiral molecule in a uniform external magnetic field. In our model
chirality breaks the parity symmetry and the magnetic field breaks time reversal, so
it is symmetry allowed to have a non-vanishing value for the momentum of the electromagnetic field.
In this case no classical contribution to the Abraham force exists since there is no external electric field. In this article we perform a
semiclassical computation of the Casimir momentum for the same model and compare it with the quantum result. We explain the physical mechanisms which mediate the transfer
of linear momentum from the EM vacuum to the chiral molecule and we analyse the transfer of energy.\\
\indent The article is organized as follows. In Section \ref{sec2} we present the model. Next, we perform a semiclassical computation of the Casimir momentum in Section \ref{sec3}. In Section
\ref{Qapproach} we review the quantum computation of Ref.\cite{PRLDonaire}, paying special attention to the physical mechanisms involved in the transfer of momentum.
In Section \ref{sec5} we explain the exchange of energy between molecule, quantum vacuum and magnetic field. In Section \ref{sec6} we summarize our conclusions.

\section{The model}\label{sec2}

We propose the simplest model for a chiral molecule that exhibits all necessary features to leading order in perturbation theory:
broken mirror symmetry, Zeeman splitting of energy levels and coupling to the
quantum vacuum, and we neglect relativistic effects. In our model the optical activity of the molecule is determined by a single chromophoric electron within a
 chiral object which is further simplified to be a two-particle system in which the chromophoric electron of
charge $q_{e}=-e$ and mass $m_{e}$ is bound to a nucleus of effective charge $q_{N}=e$ and mass $m_{N}\gg m_{e}$. The binding interaction is modeled by a
harmonic oscillator potential, $V^{HO}=\frac{\mu}{2}(\omega_{x}^{2}x^{2}+\omega_{y}^{2}y^{2}+\omega_{z}^{2}z^{2})$,
to which we add a term $V_{C}=C\:xyz$ to break the mirror symmetry perturbatively in first order. The coordinates $x$, $y$, $z$ are those of the relative position vector,
$\mathbf{r}=\mathbf{r}_{N}-\mathbf{r}_{e}$, and $\mu=\frac{m_{N}m_{e}}{M}$ with $M=m_{N}+m_{e}$. The center of mass position vector is
$\mathbf{R}=(m_{N}\mathbf{r}_{N}+m_{e}\mathbf{r}_{e})/M$. The conjugate momentum of $\mathbf{r}$ is 
$\mathbf{p}=\mu(\mathbf{p}_{N}/m_{N}-\mathbf{p}_{e}/m_{e})$, while the conjugate momentum of $\mathbf{R}$ is the total conjugate momentum,
$\mathbf{P}=\mathbf{p}_{e}+\mathbf{p}_{N}$. $V_{C}$
was first introduced by Condon \emph{et al}. \cite{Condon1,Condon2} to explain the rotatory power of chiral compounds with a single oscillator model. Both the
anisotropy in $V^{HO}$ and the chiral potential $V_{C}$ are determined by the Coulomb interaction of the two-body system with the rest of atoms within the molecule.
In particular, the parameter $C$ is the sum of all third order coefficients of the expansion of
the Coulomb interaction of the chromophoric group with the surrounding charges around their mean distance [see Eq.(42) of Ref.\cite{Condon2}].
It is a  pseudo-scalar which does not necessarily vanish for a chiral environment. In principle, all parameters of this model can be computed
\emph{ab initio} from the experimental data \cite{Condon2}.\\
\indent When an external uniform and constant magnetic field
$\mathbf{B}_{0}$ is applied the total Hamiltonian of the system reads, $H=H_{0}+H_{EM}+W$, with
\begin{align}
H_{0}&=\sum_{i=e,N}\frac{1}{2m_{i}}[\mathbf{p}_{i}-q_{i}\mathbf{A}_{0}(\mathbf{r}_{i})]^{2}+V^{HO}+V_{C},\label{H0}\\
H_{EM}&=\sum_{\mathbf{k},\mathbf{\epsilon}}\hbar\omega_{\mathbf{k}}(a^{\dagger}_{\mathbf{k}\mathbf{\epsilon}}a_{\mathbf{k}\mathbf{\epsilon}}+\frac{1}{2})
+\frac{1}{2\mu_{0}}\int\textrm{d}^{3}r\mathbf{B}_{0}^{2},\label{HEM}\\
W&=\sum_{i=e,N}\frac{-q_{i}}{m_{i}}[\mathbf{p}_{i}-q_{i}\mathbf{A}_{0}(\mathbf{r}_{i})]\cdot\mathbf{A}(\mathbf{r}_{i})
 +\frac{q^{2}_{i}}{2m_{i}}\mathbf{A}^{2}(\mathbf{r}_{i}),\label{elW}
\end{align}
where $W$ is the minimal coupling interaction potential.
In the vector potential we have separated the contribution of the external classical field, $\mathbf{A}_{0}(\mathbf{r}_{i})=\frac{1}{2}\mathbf{B}_{0}\wedge\mathbf{r}_{i}$,
from the one of the quantum field operator, $\mathbf{A}(\mathbf{r}_{i})$. Note that, having incorporated the internal electrostatic interaction of the two-body
system within $V^{HO}$ and the electrostatic interaction of the system with the surrounding within $V_{C}$, the EM vector potential in $W$ is just transverse.
In the sum, $a^{\dagger}_{\mathbf{k}\mathbf{\epsilon}}$ and $a_{\mathbf{k}\mathbf{\epsilon}}$
are the creation and annihilation operators of photons with momentum
$\hbar\mathbf{k}$, frequency $\omega_{\mathbf{k}}=ck$ and polarization vector $\mathbf{\epsilon}$ respectively.
The magnetostatic energy is a constant irrelevant to us that we will discard.\\
\indent In the absence of coupling to the vacuum field, the system with Hamiltonian $H_{0}$ possesses a conserved pseudo-momentum,
$\mathbf{K}_{0}=\mathbf{P}_{\textrm{kin}}+e\mathbf{B}_{0}\wedge\mathbf{r}$, which satisfies $[H_{0},\mathbf{K}_{0}]=\mathbf{0}$ and has continuous eigenvalues
 $\mathbf{Q}$. Here $\mathbf{P}_{\textrm{kin}}$ is the kinetic momentum of the center of mass, $\mathbf{P}_{\mathrm{kin}}=M\dot{\mathbf{R}}$, which relates to $\mathbf{P}$ 
 through $\mathbf{P}_{\textrm{kin}}=\mathbf{P}-\frac{e}{2}\mathbf{B}_{0}\wedge\mathbf{r}-e[\mathbf{A}(\mathbf{r}_{N})-\mathbf{A}(\mathbf{r}_{e})]$. 
The unitary operator U$=\exp{[i(\mathbf{Q}-\frac{e}{2}\mathbf{B}_{0}\wedge\mathbf{r})\cdot\mathbf{R}/\hbar]}$ maps the Hamiltonian $H_{0}$ into
$\tilde{H}_{0}=$U$^{\dagger}H_{0}$U, which conveniently separates the motion of the center of mass from the relative motion \cite{Herold,Dippel},
\begin{equation}
\tilde{H}_{0}=\frac{1}{2M}\mathbf{Q}^{2}+\frac{1}{2\mu}\mathbf{p}^{2}+V^{HO}+V_{C}+V_{Z}+\Delta V.\label{effective}
\end{equation}
In this equation $V_{Z}=\frac{e}{2\mu^{*}}(\mathbf{r}\wedge\mathbf{p})\cdot\mathbf{B}_{0}$ is the Zeeman potential
with $\mu^{*}=\frac{m_{N}m_{e}}{m_{N}-m_{e}}$. Terms of order $\textrm{Q}\textrm{B}_{0}$ and $\textrm{B}_{0}^{2}$ are cast in
$\Delta V=(e^{2}/2)(1/M+\mu/\mu^{*2})(\mathbf{r}\wedge\mathbf{B}_{0})^{2}+e\mathbf{Q}\cdot(\mathbf{r}\wedge\mathbf{B}_{0})/M$. In the following,
both $V_{C}$ and $V_{Z}$ will be considered first order perturbations to the harmonic oscillator potential and higher order perturbative terms like
those in $\Delta V$ will be neglected.\\
\indent The ground state of the Hamiltonian $\tilde{H}_{0}$ is, up to order $C$B$_{0}$ in stationary perturbation theory,
\begin{align}
|\tilde{\Omega}_{0}\rangle&=|0\rangle-\mathcal{C}|111\rangle
-i\mathcal{B}_{0}^{z}\eta^{yx}|110\rangle\nonumber\\
&+i\mathcal{B}_{0}^{z}\mathcal{C}\eta^{yx}\left(|001\rangle+2|221\rangle\right)\nonumber\\
&-\sqrt{2}i\mathcal{B}_{0}^{z}\mathcal{C}\left(\frac{2\omega_{x}-\omega_{z}\eta^{yx}}{\omega_{z}+2\omega_{x}}|201\rangle
-\frac{2\omega_{y}+\omega_{z}\eta^{yx}}{\omega_{z}+2\omega_{y}}|021\rangle\right)\nonumber\\
&+\sum\textrm{ cyclic permutations}.\label{vacuum}
\end{align}
Correspondingly, the ground state of $H_{0}$ is $|\Omega_{0}\rangle=$U$|\tilde{\Omega}_{0}\rangle$, with a pseudo-momentum $\mathbf{Q}_{0}$ to be fixed.
The fact that $\langle\Omega_{0}|\mathbf{r}|\Omega_{0}\rangle=\mathbf{0}$ implies that $\mathbf{Q}_{0}$ is the kinetic momentum of the oscillator,
i.e., no other contribution exists to the pseudo-momentum $\langle\mathbf{K}_{0}\rangle$ of the bare molecule in its ground state.
In the above equation the states $|n_{x} n_{y} n_{z}\rangle$ refer to the eigenstates of the harmonic oscillator Hamiltonian. The dimensionless parameters are, $\mathcal{B}_{0}^{i}=\frac{eB_{0}^{i}}{4\mu^{*}\sqrt{\omega_{j}\omega_{k}}}$ with
$i\neq j\neq k$ and $i\neq k$, $\mathcal{C}=\frac{C\hbar^{1/2}}{(2\mu)^{3/2}(\omega_{x}+\omega_{y}+\omega_{z})(\omega_{x}\omega_{y}\omega_{z})^{1/2}}$,
$\eta^{ij}=\frac{\omega_{i}-\omega_{j}}{\omega_{i}+\omega_{j}}$. Here, the indices $i,j,k$ take on the three spatial directions, $x,y,z$.
The $\eta$ factors are assumed to be small quantities which quantify the anisotropy of the oscillator.
They \emph{all} have to be nonzero for the optical activity of the molecule to survive rotational averaging. The dimensionless parameters,
$\mathcal{C}$, $\mathcal{B}_{0}^{i}$ and $\eta^{ij}$, are the expansion
parameters of our perturbative calculations. We will restrict to
the lowest order in them all.

\section{Semiclassical approach}\label{sec3}
In Ref.\cite{vTg2008} van Tiggelen has derived an expression for the Casimir momentum using a Green's function formalism which is compatible with Nelson's
prescription. That formalism was latter applied in Ref.\cite{BabingtonvTg2010} to compute the Casimir momentum in a configuration of magnetic dipoles disposed in a
twisted H configuration.\\
\indent In our case we are interested in the Casimir momentum in the presence of a single molecule. As noticed in Ref.\cite{Rikken1}, the Casimir momentum which derives from Abraham and Nelson's
prescriptions coincide approximately for a diluted medium
(same applies to the forces generated by time varying fields). For the case of a single particle the so-called Abraham momentum reads $\mathbf{d}\wedge\mathbf{B}$, $\mathbf{d}$ being
the electric dipole moment of the molecule. Here
we apply a linear response formalism upon $\mathbf{d}\wedge\mathbf{B}$
following a similar semiclassical treatment to the one for the Lamb shift of a single dipole or the one for the van der Waals forces between atoms \cite{MilonniPhysScripta,Craigbook,CraigThiru}.
This is  equivalent
to the Green's function formalism of Refs.\cite{vTg2008,BabingtonvTg2010} considering a single magnetochiral scatterer.\\
\indent We aim to compute the
vacuum expectation value $\mathbf{P}^{\textrm{Cas}}=\Re{\{\langle\mathbf{d}\wedge\mathbf{B}\rangle\}}$. We start the calculation by decomposing both the electric dipole
operator and the magnetic field into induced (\emph{ind}) and free (\emph{fr}) components. The induced operators relate to the free operators through the linear
susceptibilities, i.e., polarizabilities and Green's functions. Any space-time dependent free operator $X_{fr}(\mathbf{R};t)$ can be written in Fourier space as a sum of positive frequency
and negative frequency components,
$X_{fr}(\mathbf{R};t)=\int_{0}^{\infty}\textrm{d}\omega[X_{fr}(\mathbf{R};\omega)e^{-i\omega t}
+X_{fr}^{\dagger}(\mathbf{R};\omega)e^{i\omega t}]$,
where $X_{fr}^{\dagger}(\mathbf{R};\omega)$ and $X_{fr}(\mathbf{R};\omega)$ are the $\omega$-mode creation and annihilation operators of $X_{fr}$ respectively. In the following we will be interested in the EM field and dipole moment operators evaluated at the position of the molecule, $\mathbf{R}_{0}$. Therefore, unless necessary, we will omit the dependence of the operators on the position. In the case of the electric field of a plane wave  propagating in free space in the direction $\hat{\mathbf{k}}$, we find   
$\mathbf{E}_{fr}(\mathbf{R};\omega)\propto e^{i\mathbf{k}\cdot\mathbf{R}}$, with $\mathbf{k}=\omega\hat{\mathbf{k}}/c$, and likewise for the magnetic field. Therefore, together with the dependence on $\omega$ we will add the dependence on $\mathbf{k}$ to the electric field annihilation/creations operator of a plane wave, $\mathbf{E}_{fr}(\mathbf{k},\omega)$ --\emph{idem} for the magnetic field.\\
\indent Since the relation between induced and free operators is linear and the action of annihilation operators on the vacuum vanishes, it follows from $\mathbf{P}^{\textrm{Cas}}=\Re{\{\langle\mathbf{d}\wedge\mathbf{B}\rangle\}}$ that
\begin{equation}
\mathbf{P}^{\textrm{Cas}}=\Re{\{\langle\mathbf{d}_{fr}\wedge\mathbf{B}_{ind}\rangle\}}+\Re{\{\langle\mathbf{d}_{ind}\wedge\mathbf{B}_{fr}\rangle\}}.\label{PCASO}
\end{equation}
In Ref.\cite{EPJDonaire} we have obtained the constitutive equations of our magneto-chiral model by computing the response of the molecule to a 
 monochromatic EM plane wave of frequency $\omega$ and wave vector $\mathbf{k}$. Up to electric
quadrupole contributions, the induced electric dipole moment operator reads,
\begin{align}
\textrm{d}^{ind}_{i}(\omega)&=\alpha_{E}\delta_{ij}\textrm{E}_{fr}^{j}(\mathbf{k},\omega) +\chi\epsilon_{ijk}\textrm{B}^{j}_{0}\dot{\textrm{E}}_{fr}^{k}(\mathbf{k},\omega)
\nonumber\\&-\beta\delta_{ij}\dot{\textrm{B}}^{j}_{fr}(\mathbf{k},\omega)+\gamma\epsilon_{ijk}\textrm{B}^{j}_{0}\textrm{B}_{fr}^{k}(\mathbf{k},\omega)\nonumber\\&+
\frac{1}{2}\xi[(\mathbf{B}_{0}\cdot\mathbf{k})\textrm{E}^{fr}_{i}(\mathbf{k},\omega)+(\mathbf{B}_{0}\cdot\mathbf{E}_{fr}(\mathbf{k},\omega))\textrm{k}_{i}],\label{dind}
\end{align}
where a rotational average is implicit in this equation. The factor $\alpha_{E}$ is the ordinary electric polarizability,
$\chi$ describes the Faraday
effect, $\beta$ is the molecular rotatory factor responsible for
the natural optical activity, and $\gamma$ and $\xi$  give rise to the magnetochiral
anisotropy. The expressions of all these factors have been summarized in \ref{appendB}. It has been shown in Ref.\cite{EPJDonaire} that, once the rotational
power and the refractive index of a compound are given --i.e., the parameters $\alpha_{E}$ and $\beta$, all other parameters of our model can be deduced.\\
\indent Defining the effective electric polarizability and the crossed magneto-electric polarizability tensors respectively as,
\begin{align}
\alpha_{EE}^{ij}(\mathbf{k},\omega)&=[\alpha_{E}+\frac{1}{2}\xi(\mathbf{B}_{0}\cdot\mathbf{k})]\delta^{ij}+\textrm{k}^{i}\textrm{B}_{0}^{j}-i\omega\chi\epsilon_{ikj}\textrm{B}^{k}_{0},
\nonumber\\
\alpha_{EM}^{ij}(\omega)&=i\omega\beta\delta^{ij}+\gamma\epsilon_{ikj}\textrm{B}^{k}_{0},
\end{align}
we can write Eq.(\ref{dind}) as,
\begin{equation}
\textrm{d}^{ind}_{i}(\omega)=\alpha_{EE}^{ij}(\mathbf{k},\omega)\textrm{E}^{fr}_{j}(\mathbf{k},\omega)+\alpha_{EM}^{ij}(\omega)\textrm{B}^{fr}_{j}(\mathbf{k},\omega).\label{dresp}
\end{equation}
Using Maxwell's equation for the free fields, $\dot{\mathbf{E}}_{fr}=\mathbf{\nabla}\wedge\mathbf{B}_{fr}$, we can write $\mathbf{d}_{ind}(\omega)$ as a response
to the free electric field alone,
\begin{align}
\mathbf{d}^{ind}(\omega)&=\alpha_{E}\mathbf{E}_{fr}(\mathbf{k},\omega)-i\omega\chi\mathbf{B}_{0}\wedge\mathbf{E}_{fr}(\mathbf{k},\omega)
+i\beta\mathbf{k}\wedge\mathbf{E}_{fr}(\mathbf{k},\omega)\nonumber\\&+(\xi/2-\gamma/\omega)(\mathbf{B}_{0}\cdot\mathbf{k})\mathbf{E}_{fr}(\mathbf{k},\omega)
\nonumber\\&+(\xi/2+\gamma/\omega)[\mathbf{B}_{0}\cdot\mathbf{E}_{fr}(\mathbf{k},\omega)]\mathbf{k}.
\end{align}
From here it is obvious that the non-reciprocal response comes from the fourth term on the r.h.s., which depends on the relative direction of the wave vector, $\mathbf{k}$, with respect
to the external magnetic field. We will denote this non-reciprocal (\emph{nr}) polarizability by
$\alpha_{nr}(\mathbf{k},\omega)=(\xi/2-\gamma/\omega)(\mathbf{B}_{0}\cdot\mathbf{k})$. It is at the origin of the magnetochiral birefringence \cite{EPJDonaire}.
The induced magnetic field can be written as a linear response to the free dipole located at $\mathbf{R}_{0}$,
\begin{align}
\mathbf{B}_{ind}(\mathbf{R},\omega)&=\epsilon_{0}^{-1}c^{-1}[\mathbb{G}^{(0)}_{me}(\mathbf{R},\mathbf{R}_{0};\omega)\cdot\mathbf{d}_{fr}(\mathbf{R}_{0},\omega)
\nonumber\\&+c^{-1}\mathbb{G}^{(0)}_{mm}(\mathbf{R},\mathbf{R}_{0};\omega)\cdot\mathbf{m}_{fr}(\mathbf{R}_{0},\omega)],\label{bind}
\end{align}
where $\mathbf{m}_{fr}(\mathbf{R}_{0},\omega)$ is the $\omega$-mode of the free magnetic dipole moment operator and the Green functions $\mathbb{G}^{(0)}_{mm}$ and
$\mathbb{G}^{(0)}_{me}$  relates to the Green function of Maxwell's equation for the EM vector potential in free space,
\begin{equation}\label{Maxwellb}
\Bigl[\omega^{2}\mathbb{I}-\mathbf{\nabla}\wedge\mathbf{\nabla}\wedge\Bigr]\mathbb{G}^{(0)}(\mathbf{R}-\mathbf{R}';\omega)
=\delta^{(3)}(\mathbf{R}-\mathbf{R}')\mathbb{I},
\end{equation}
\begin{align}
&\textrm{through }\qquad\quad
\mathbb{G}^{(0)}_{mm}(\mathbf{R},\mathbf{R}';\omega)=\mathbf{\nabla}_{\mathbf{R}}\wedge\mathbb{G}^{(0)}(\mathbf{R},\mathbf{R}';\omega)\wedge\mathbf{\nabla}_{\mathbf{R}'},
\nonumber\\&\textrm{and }\qquad\quad
\mathbb{G}^{(0)}_{me}(\mathbf{R},\mathbf{R}';\omega)=ik\mathbf{\nabla}_{\mathbf{R}}\wedge\mathbb{G}^{(0)}(\mathbf{R},\mathbf{R}';\omega).\nonumber
\end{align}
\indent When substituting the induced operators in Eq.(\ref{PCASO}) as functions of the free operators we find the vacuum expectation values of bilinear operators.
Those expectation values (dipole and field quadratic fluctuations) relate to the imaginary part of their respective linear response functions through
the fluctuation-dissipation theorem \cite{Butcher},
\begin{align}
\langle \mathbf{B}_{fr}(\mathbf{R},\omega)\otimes\mathbf{E}^{\dagger}_{fr}(\mathbf{R}',\omega')\rangle&=
\frac{\hbar}{\pi\epsilon_{0}c}\Im{\{\mathbb{G}^{(0)}_{me}(\mathbf{R},\mathbf{R}';\omega)\}}\delta(\omega-\omega'),\nonumber\\
\langle \mathbf{B}_{fr}(\mathbf{R},\omega)\otimes\mathbf{B}_{fr}^{\dagger}(\mathbf{R}',\omega')\rangle&=
\frac{\hbar}{\pi\epsilon_{0}c^{2}}\Im{\{\mathbb{G}^{(0)}_{mm}(\mathbf{R},\mathbf{R}';\omega)\}}\delta(\omega-\omega'),\nonumber
\end{align}
\begin{align}
\langle\mathbf{d}_{fr}(\omega)\otimes\mathbf{d}^{\dagger}_{fr}(\omega')\rangle&=\frac{\hbar}{\pi}
\Im{\{\alpha_{EE}\}}\delta(\omega-\omega'),\nonumber\\
\langle d^{fr}_{i}(\omega)\otimes m^{fr\dagger}_{j}(\omega')\rangle&=
\frac{\hbar}{2\pi i}[\alpha_{EM}-\alpha_{ME}^{\dagger}]_{ij}\delta(\omega-\omega')\nonumber\\&=
\frac{\hbar}{\pi}\epsilon_{ikj}\textrm{B}^{k}_{0}\Im{\{\gamma\}}\delta(\omega-\omega')\nonumber\\&+\textrm{reciprocal terms}\label{13},
\end{align}
where it is understood that the dipole moment operators act upon the location of the only molecule at $\mathbf{R}_{0}$. In the last equation above we use the result
obtained in Ref.\cite{EPJDonaire}, $\alpha_{EM}=-\alpha_{ME}$, and we restrict to nonreciprocal
terms for simplicity. Using the linear response relations
of Eqs.(\ref{dresp}) and (\ref{bind}) in Eq.(\ref{PCASO})  and applying the expectation values of Eq.(\ref{13}) we end up with the relation,
\begin{align}
P^{\textrm{Cas}}_{i}&=\frac{\hbar}{\pi\epsilon_{0}c}\varepsilon_{ij}^{\:\:\:k}\Im\int_{0}^{\infty}\textrm{d}\omega\int\frac{\textrm{d}^{3}q}{(2\pi)^{3}}
\big[\alpha_{EE}^{jp}(\mathbf{q},\omega)\tilde{G}_{em,pk}^{(0)}(\mathbf{q},\omega)\nonumber\\
&+c^{-1}\alpha_{EM}^{jp}(\omega)\tilde{G}_{mm,pk}^{(0)}(\mathbf{q},\omega)\big],\label{cc}
\end{align}
where $\tilde{\mathbb{G}}_{em}^{(0)}(\mathbf{q},\omega)$ and $\tilde{\mathbb{G}}_{mm}^{(0)}(\mathbf{q},\omega)$ are the Fourier transforms of $\mathbb{G}_{em}^{(0)}(\mathbf{R},\mathbf{R}^{'};\omega)$ and $\mathbb{G}_{mm}^{(0)}(\mathbf{R},\mathbf{R}^{'};\omega)$  in $\mathbf{q}$-space respectively, 
\begin{eqnarray}
\tilde{G}_{em,pk}^{(0)}(\mathbf{q},\omega)&=&\frac{-k}{k^{2}-q^{2}+i\mu}\varepsilon_{pkr}q^{r},\nonumber\\
\tilde{G}_{mm,pk}^{(0)}(\mathbf{q},\omega)&=&\frac{-q^{2}}{k^{2}-q^{2}+i\mu}(\delta_{pk}-q_{p}q_{k}/q^{2}),\:\mu\rightarrow0^{+},
\end{eqnarray}
which are necessary to account for the dependence of $\alpha_{EE}$ on the wave vector. Symmetry considerations imply that only the nonreciprocal terms of $\alpha_{EE}$ and $\alpha_{EM}$ in $\alpha_{nr}$ survive the angular integration. After integrating  over $\mathbf{q}$ we arrive at
\begin{equation}
\mathbf{P}^{\textrm{Cas}}=\frac{\hbar\mathbf{B}_{0}}{6\pi^{2}\epsilon_{0}c^{5}}\Re\int_{0}^{\infty}\textrm{d}\omega\:\omega^{4}(\xi/2-\gamma/\omega).\label{ec1}
\end{equation}
This expression is nothing but Feigel's formula \cite{Feigel} simplified to the case of a diluted medium in the one-particle limit $\mathcal{V}\rho\rightarrow1$, $\mathcal{V}$
being the total volume and $\rho$ the numerical density of molecules \cite{MilonniPhysScripta}. Hence, following Croze's argument \cite{Croze} we can write the above integral as
the sum over the momenta of 'dressed' normal modes, $n\hbar\mathbf{k}$, with $n$ being the index of refraction,
\begin{equation}
\mathbf{P}^{\textrm{Cas}}=\mathcal{V}\sum_{\mathbf{k},\mathbf{\epsilon}} n\hbar\mathbf{k}=\mathcal{V}\sum_{\mathbf{k},\mathbf{\epsilon}} \hbar\delta n_{MCh}\mathbf{k},\label{ec2}
\end{equation}
with $\delta n_{MCh}$ the nonreciprocal part of the refractive index due to magnetochiral birefringence.  At leading order in $\rho$ it reads \cite{EPJDonaire},
$\delta n_{MCh}=\rho(\xi/2-\gamma/\omega)(\mathbf{B}_{0}\cdot\mathbf{k})/\epsilon_{0}$. In the one-particle limit, $\mathcal{V}\rho\rightarrow1$, and turning the summation
into an integral in the continuum limit, Eqs.(\ref{ec1}) and
(\ref{ec2}) coincide. From the semiclassical result it follows that
$\mathbf{P}^{\textrm{Cas}}$ is the momentum of radiative modes only. This result would reinforce the interpretation that the Casimir momentum originates in the non-reciprocity of the spectrum of normal modes
propagating in the effective medium, i.e. $\tilde{\omega}(\mathbf{k})\neq\tilde{\omega}(-\mathbf{k})$, with $\tilde{\omega}(\pm\mathbf{k})$ the 'effective' frequencies of photons
propagating in opposite directions with bare momentum vectors $\pm\hbar\mathbf{k}$.
\indent To conclude, we use the explicit formulas for $\gamma$ and $\xi$ in \ref{appendB} and integrate over frequencies,
\begin{equation}\label{PCascalss}
\mathbf{P}^{\textrm{Cas}}=-\frac{\hbar^{2}e^{3}C\mathbf{B}_{0}}{1458\pi^{2}c^{5}\epsilon_{0}\omega_{0}\mu^{2}\mu^{*2}}\eta^{zy}\eta^{yx}\eta^{xz},
\end{equation}
where $\omega_{0}=(\omega_{x}+\omega_{y}+\omega_{z})/3$.
It is remarkable that the frequency integral does not present any UV divergence, unlike the expression for $\mathbf{P}^{\textrm{Cas}}$ in Ref.\cite{Feigel} for
a magneto-electric medium in crossed fields. We will see later on that this semiclassical result is included in the microscopic QED result,
but it enters as a small correction to the leading term.


\section{Quantum approach}\label{Qapproach}
The system of the Hamiltonian $H$ possesses a conserved pseudo-momentum \cite{Kawka1,Herold,Dippel},
\begin{equation}\label{Po}
\mathbf{K}=\mathbf{P}+\frac{e}{2}\mathbf{B}_{0}\wedge\mathbf{r}+\sum_{\mathbf{k},\mathbf{\epsilon}}\hbar\mathbf{k}(a^{\dagger}_{\mathbf{k}\mathbf{\epsilon}}a_{\mathbf{k}\mathbf{\epsilon}}+\frac{1}{2}),
\end{equation}
which satisfies $[H,\mathbf{K}]=\mathbf{0}$. Its eigenvalues are therefore good quantum numbers. 
The terms in $\mathbf{K}$ can be also arranged as,
\begin{equation}\label{P}
\mathbf{K}=\mathbf{P}_{\textrm{kin}}+\mathbf{P}_{\textrm{Abr}}+\mathbf{P}^{\textrm{Cas}}_{\parallel}+\mathbf{P}^{\textrm{Cas}}_{\perp},
\end{equation}
where, beside the kinetic momentum, $\mathbf{P}_{\textrm{Abr}}=e\mathbf{B}_{0}\wedge\mathbf{r}$ is the Abraham momentum and we define the Casimir momentum operator,
$\mathbf{P}^{\textrm{Cas}}=\mathbf{P}^{\textrm{Cas}}_{\parallel}+\mathbf{P}^{\textrm{Cas}}_{\perp}$, as the
momentum operator of the vacuum field. $\mathbf{P}^{\textrm{Cas}}$ is composed of a longitudinal Casimir momentum,
$\mathbf{P}^{\textrm{Cas}}_{\parallel}=e[\mathbf{A}(\mathbf{r}_{N})-\mathbf{A}(\mathbf{r}_{e})]$, and a transverse Casimir momentum,
$\mathbf{P}^{\textrm{Cas}}_{\perp}=\sum_{\mathbf{k},\mathbf{\epsilon}}\hbar\mathbf{k}(a^{\dagger}_{\mathbf{k}\mathbf{\epsilon}}a_{\mathbf{k}\mathbf{\epsilon}}+\frac{1}{2})$
\cite{Cohen}. The latter is just the sum of the momenta $\hbar \mathbf{k}$ of
radiative photons. The longitudinal momentum is more subtle. It stems from the transverse electromagnetic gauge field coupled to electric charges.
Note that in the Coulomb gauge $\mathbf{A}$ is fully transverse. $\mathbf{P}^{\textrm{Cas}}_{\parallel}$ is referred to as "longitudinal" component since it can be written as the integral
over the vector product of the longitudinal Coulomb electric field and
the magnetic field generated by the charges, $\epsilon_{0}\int\textrm{d}^{3}r\mathbf{E}^{Coul}\wedge\mathbf{B}$ \cite{Cohen}.\\
\indent Under the action of a time-varying magnetic field, $\mathbf{B}_{0}(t)$, the time derivative of the expectation value of $\mathbf{K}$
in the ground state, $\langle\mathbf{K}\rangle$, vanishes,
\begin{equation}
\frac{\textrm{d}\langle\mathbf{K}\rangle}{\textrm{d}t}=i\hbar^{-1}\langle[H,\mathbf{K}]\rangle+e\frac{\partial\mathbf{B}_{0}(t)}{\partial t}\wedge\langle\mathbf{r}\rangle=\mathbf{0}.\nonumber
\end{equation}
This follows from $[H,\mathbf{K}]=\mathbf{0}$ and $\langle\mathbf{r}\rangle=\mathbf{0}$ for the chiral but unpolarized ground state
\footnote{The oscillator in motion generates a dipole moment $\alpha_{E}(0)\mathbf{Q}_{0}\wedge\mathbf{B}_{0}/M$ \cite{Herold,Dippel}.}.
The latter ensures also that the variation of the Abraham momentum vanishes in the ground state. From the expectation value of Eq.(\ref{P}) it follows that,
for an arbitrary variation of the magnetic field, the variation of the kinetic momentum of the chiral oscillator is equivalent in magnitude and opposite
in sign to the variation of the Casimir momentum of the vacuum field,
\begin{equation}\label{dPs}
\delta\langle\mathbf{P}_{\textrm{kin}}\rangle=-\delta\langle\mathbf{P}^{\textrm{Cas}}\rangle.
\end{equation}
\indent Let us consider the molecule initially at rest in its ground state at zero magnetic field. During the switching of the magnetic field
Eq.(\ref{dPs}) implies that $\langle\mathbf{P}_{\textrm{kin}}\rangle=-\langle\mathbf{P}^{\textrm{Cas}}\rangle$ at any time. This relation makes
the Casimir momentum an observable quantity. In particular this equality
holds well after the switching has ended and the magnetic field achieves a stationary value $\mathbf{B}_{0}$. In the following we
evaluate the Casimir momentum in the asymptotically stationary situation in which the molecule is in its ground state at constant kinetic momentum, $\mathbf{Q}_{0}$,
constant magnetic field, $\mathbf{B}_{0}$ and once coupled to the EM vacuum. We denote by $|\Omega\rangle$ this asymptotic state of the molecule, which
we will compute applying up to second order perturbation
theory to $|\Omega_{0}\rangle$ with the interaction potential $W$. We define $\langle\mathbf{P}^{\textrm{Cas}}\rangle=\langle\Omega|\mathbf{P}^{\textrm{Cas}}|\Omega\rangle$,
and calculate separately the transverse momentum,
$\langle\mathbf{P}_{\perp}^{\textrm{Cas}}\rangle=\sum_{\mathbf{k},\mathbf{\epsilon}}\hbar\mathbf{k}\langle\Omega|a^{\dagger}_{\mathbf{k}\mathbf{\epsilon}}a_{\mathbf{k}\mathbf{\epsilon}}|\Omega\rangle$,
and the longitudinal momentum,
$\langle\mathbf{P}_{\parallel}^{\textrm{Cas}}\rangle=e\langle\Omega|\mathbf{A}(\mathbf{r}_{N})-\mathbf{A}(\mathbf{r}_{e})|\Omega\rangle$.\\
\subsection{Transverse Casimir momentum}
\begin{figure}[htb]\label{fig1}
\includegraphics[height=9.7cm,width=9.0cm,clip]{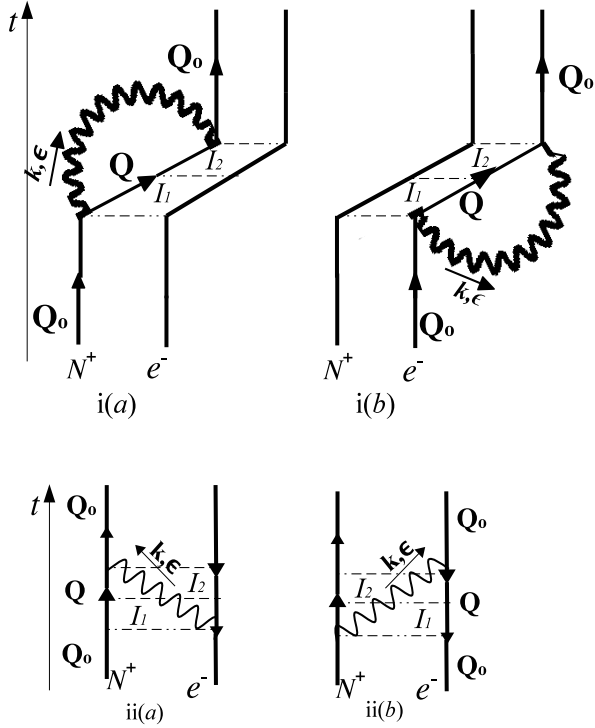}
\caption{i.Feynman diagrams of the dominant processes contributing to $\langle\mathbf{P}^{\textrm{Cas}}\rangle$. For the case of $\langle\mathbf{P}^{\textrm{Cas}}_{\perp}\rangle$
there are two intermediate atomic states, $I$ and $I'$, and diagram i$(a)$ is negligible. For the case of $\langle\mathbf{P}^{\textrm{Cas}}_{\parallel}\rangle$
both diagrams contribute at the same order, and there is only one intermediate atomic state.\\
ii.Feynman diagrams of the subdominant processes contributing to $\langle\mathbf{P}^{\textrm{Cas}}\rangle$. Virtual photons of momentum $\mathbf{k}$ and polarization $\mathbf{\epsilon}$ fly between the electron and the
nucleus, yielding an effective wavelength cut-off of the order of the interparticle distance, $\sim\sqrt{\hbar/\mu\omega_{0}}$.}
\end{figure}

For the computation of $\langle\mathbf{P}_{\perp}^{\textrm{Cas}}\rangle$ at $\mathcal{O}(\mathcal{C}\mathcal{B}_{0})$ and lowest order in the fine structure constant,
$\alpha=e^{2}/4\pi\epsilon_{0}\hbar c$, we need to
compute $|\Omega\rangle$ applying second order perturbation theory to $|\Omega_{0}\rangle$. Note that this implies applying up to fourth-order perturbation theory
in $V_{Z}+V_{C}+W$ to the ground state of the harmonic oscillator Hamiltonian, $|0\rangle$. We use the U-transformed states and the U-transformed potential, with U$=\exp{[-i\frac{e}{2\hbar}(\mathbf{B}_{0}\wedge\mathbf{r})\cdot\mathbf{R}]}$,
\begin{eqnarray}\label{tW}
\tilde{W}&=&-\frac{e}{m_{N}}\left(\mathbf{p}+\frac{m_{N}}{M} \mathbf{P}-\frac{e}{2}\mathbf{B}_{0}\wedge\mathbf{r}\right)\cdot
\mathbf{A}(\mathbf{R}+\frac{m_{e}}{M}\mathbf{r})\nonumber\\
&-&\frac{e}{m_{e}}\left(\mathbf{p}-\frac{m_{e}}{M} \mathbf{P}+\frac{e}{2}\mathbf{B}_{0}\wedge\mathbf{r}\right)\cdot
\mathbf{A}(\mathbf{R}-\frac{m_{N}}{M}\mathbf{r})\nonumber\\
&+&\frac{e^{2}}{2m_{N}}\textrm{A}^{2}(\mathbf{R}+\frac{m_{e}}{M}\mathbf{r})
+\frac{e^{2}}{2m_{e}}\textrm{A}^{2}(\mathbf{R}-\frac{m_{N}}{M}\mathbf{r}),\label{tildeW}
\end{eqnarray}
to arrive at,
\begin{eqnarray}
\langle\mathbf{P}^{\textrm{Cas}}_{\perp}\rangle&=&\sum_{\mathbf{Q},I,\gamma_{\mathbf{k}\mathbf{\epsilon}}}\sum_{\mathbf{Q}',I',\gamma_{\mathbf{k}'\mathbf{\epsilon}'}}
\sum_{\mathbf{k}'',\mathbf{\epsilon}''}
\frac{\langle\mathbf{Q}_{0},\tilde{\Omega}_{0}|\tilde{W}|\mathbf{Q},I,\gamma\rangle}{\hbar^{2}Q_{0}^{2}/2M+E_{0}-E_{\mathbf{Q},I,\mathbf{k}}}\nonumber\\
&\times&\langle\mathbf{Q},I,\gamma|\hbar\mathbf{k}''a^{\dagger}_{\mathbf{k}''\mathbf{\epsilon}''}a_{\mathbf{k}''\mathbf{\epsilon}''}
|\mathbf{Q}',I',\gamma'\rangle\nonumber\\
&\times&\frac{\langle\mathbf{Q}',I',\gamma'|\tilde{W}|\mathbf{Q}_{0},\tilde{\Omega}_{0}\rangle}{\hbar^{2}Q_{0}^{2}/2M+E_{0}-E_{\mathbf{Q}',I',\mathbf{k}'}},\label{PCos}
\end{eqnarray}
where  $|\mathbf{Q}_{0},\tilde{\Omega}_{0}\rangle=\exp{(i\mathbf{Q}_{0}\cdot\mathbf{R}/\hbar)}|\tilde{\Omega}_{0}\rangle$,
$E_{0}=\hbar(\omega_{x}+\omega_{y}+\omega_{z})/2\equiv\hbar\omega_{0}/2$ and $E_{Q,I,k},E_{Q',I',k'}$ are
the energies of the intermediate states, $|\mathbf{Q},I,\gamma\rangle=|\mathbf{Q},I\rangle\otimes|\gamma_{\mathbf{k}\mathbf{\epsilon}}\rangle$ (\emph{idem}
for the prime states). The atomic states $|\mathbf{Q},I\rangle$ are eigenstates of $\tilde{H}_{0}$ and may have a priori any pseudo-momentum $\mathbf{Q}$. The EM states,
$|\gamma_{\mathbf{k}\mathbf{\epsilon}}\rangle$, are 1-photon states with momentum $\hbar\mathbf{k}$ and polarization vector $\mathbf{\epsilon}$.
Zeros in the denominator are avoided in the summation. Writing the EM quantum field in Eq.(\ref{tildeW}) as
usual \cite{Loudon},
\begin{equation}\label{AQ}
\mathbf{A}(\mathbf{r})=\sum_{\mathbf{k},\mathbf{\epsilon}}\sqrt{\frac{\hbar}{2ck\mathcal{V}\epsilon_{0}}}
[\mathbf{\epsilon}a_{\mathbf{k}}e^{i\mathbf{k}\cdot\mathbf{r}}+\mathbf{\epsilon}^{*}a^{\dagger}_{\mathbf{k}}e^{-i\mathbf{k}\cdot\mathbf{r}}],
\end{equation}
with $\mathcal{V}$ a generic volume; passing the sums over
$\mathbf{Q}$, $\mathbf{Q}'$, $\mathbf{k}$, $\mathbf{k}'$ and
$\mathbf{k}''$ in Eq.(\ref{PCos}) to continuum integrals and by
summing over polarization states we arrive at,
\begin{align}
\langle\mathbf{P}^{\textrm{Cas}}_{\perp}\rangle&=\frac{\hbar^{2} e^{2}}{2c\epsilon_{0}}
\int\frac{\textrm{d}^{3}k\:\mathbf{k}}{(2\pi)^{3}k}\nonumber\\&\times\langle\tilde{\Omega}_{0}|\Bigl[
(\mathbf{p}/m_{e}-\mathbf{Q}_{0}/M+\frac{e}{2m_{e}}\mathbf{B}_{0}\wedge\mathbf{r})
e^{-i\frac{m_{N}}{M}\mathbf{k}\cdot\mathbf{r}}\nonumber\\&+
(\mathbf{p}/m_{N}+\mathbf{Q}_{0}/M-\frac{e}{2m_{N}}\mathbf{B}_{0}\wedge\mathbf{r})
e^{i\frac{m_{e}}{M}\mathbf{k}\cdot\mathbf{r}}\Bigr]\nonumber\\
&\times\frac{\cdot(\mathbb{I}-\frac{\mathbf{k}\otimes\mathbf{k}}{k^{2}})\cdot}
{(\hbar^{2}k^{2}/2M+\hbar ck-\hbar\mathbf{k}\cdot\mathbf{\mathbf{Q}}_{0}/M-E_{0}+H_{0})^{2}}\nonumber\\
&\times\Bigl[e^{-i\frac{m_{e}}{M}\mathbf{k}\cdot\mathbf{r}}(\mathbf{p}/m_{N}+\mathbf{Q}_{0}/M-\frac{e}{2m_{N}}\mathbf{B}_{0}\wedge\mathbf{r})\nonumber\\
&+e^{i\frac{m_{N}}{M}\mathbf{k}\cdot\mathbf{r}}(\mathbf{p}/m_{e}-\mathbf{Q}_{0}/M+\frac{e}{2m_{e}}\mathbf{B}_{0}\wedge\mathbf{r})\Bigr]|\tilde{\Omega}_{0}\rangle.
\nonumber
\end{align}
In this equation we can distinguish four terms. In two of them the exponentials to the r.h.s. and l.h.s. of the fraction compensate. They correspond to the Feynman diagrams in which a photon is created and annihilated at the position of one
of the particles, i.e., either at the nucleus or at the electron position [Fig.\ref{fig1}i($a$) and Fig.\ref{fig1}i($b$), respectively]. In the other two terms the exponentials amount to $e^{\pm i\mathbf{k}\cdot\mathbf{r}}$. They
correspond to the Feynman diagrams in which the virtual photons are created and annihilated in different particles [Fig.\ref{fig1}ii($a$) and Fig.\ref{fig1}ii($b$)]. In the latter the complex
exponentials evaluated in the states of the harmonic oscillator yield an effective cut-off for the  momentum integrals at
$k_{max}\sim\sqrt{\mu\omega_{0}/\hbar}$, making their contribution negligible w.r.t. the other diagrams. Among the diagrams of Fig.\ref{fig1}i
the dominant one is i($b$) in which virtual photons are created and annihilated at the electron position, as for the non-relativistic calculation of the Lamb shift \cite{Milonnibook}. It reads,
\begin{align}
\langle\mathbf{P}^{\textrm{Cas}}_{\perp}\rangle&=\frac{\hbar^{2} e^{2}}{2cm_{e}^{2}\epsilon_{0}}
\int\frac{\textrm{d}^{3}k\:\mathbf{k}}{(2\pi)^{3}k}\nonumber\\&\langle\tilde{\Omega}_{0}|
(\frac{m_{e}}{M}\mathbf{Q}_{0}+\mathbf{p}+\frac{e}{2}\mathbf{B}_{0}\wedge\mathbf{r})
e^{-i\frac{m_{N}}{M}\mathbf{k}\cdot\mathbf{r}}\nonumber\\
&\times\frac{\cdot(\mathbb{I}-\frac{\mathbf{k}\otimes\mathbf{k}}{k^{2}})\cdot}{(\hbar^{2}k^{2}/2M+\hbar ck-\hbar\mathbf{k}\cdot\mathbf{\mathbf{Q}}_{0}/M-E_{0}+H_{0})^{2}}\nonumber\\
&\times e^{i\frac{m_{N}}{M}\mathbf{k}\cdot\mathbf{r}}(\frac{m_{e}}{M}\mathbf{Q}_{0}+\mathbf{p}+\frac{e}{2}\mathbf{B}_{0}\wedge\mathbf{r})|\tilde{\Omega}_{0}\rangle.\label{Pperpo}
\end{align}
For the above integral not to vanish, the product of the quantum operators in the integrand must be even in $\mathbf{k}$.
We observe that when moving the complex exponential on the l.h.s. to the r.h.s. of the fraction, respecting the canonical commutation relations the
momentum $\mathbf{p}$ in $\tilde{H}_{0}$ gets shifted, $\mathbf{p}\rightarrow\mathbf{p}-\frac{m_{N}}{M}\hbar\mathbf{k}$,
\begin{align}
&e^{-i\frac{m_{N}}{M}\mathbf{k}\cdot\mathbf{r}}(\hbar^{2}k^{2}/2M+\hbar ck-\hbar\mathbf{k}\cdot\mathbf{\mathbf{Q}}_{0}/M-E_{0}+\tilde{H}_{0})^{-1}\nonumber\\
&\times e^{i\frac{m_{N}}{M}\mathbf{k}\cdot\mathbf{r}}=
[\hbar^{2}k^{2}/2m_{e}+\hbar
ck-\hbar\mathbf{k}\cdot\mathbf{p}/m_{e}\nonumber\\&-\frac{\hbar\mu}{2\mu^{*}m_{e}}(\mathbf{r}\wedge\mathbf{k})\cdot\mathbf{B}_{0}
-\hbar\mathbf{k}\cdot\mathbf{\mathbf{Q}}_{0}/M-E_{0}+\tilde{H}_{0}]^{-1}.\nonumber
\end{align}
As a result, the recoil kinetic energy in the denominators of Eq.(\ref{Pperpo})
becomes $\hbar^{2}k^{2}/2m_{e}$ and an additional term $-\hbar\mathbf{k}\cdot\mathbf{p}/m_{e}$ shows up there. Because the internal velocity of the electron
is generally much greater than the center of mass velocity, $p/m_{e}\gg Q_{0}/M$, the term  $\hbar\mathbf{k}\cdot\mathbf{\mathbf{Q}}_{0}/M$ will be a higher order correction.
It is easy to see that the contribution of all the terms linear in $\mathbf{Q}_{0}$ is of the order of $(\alpha\hbar\omega_{0}/Mc^{2})\mathbf{Q}_{0}$ \cite{Kawka2},
which we interpret as a radiative correction to the total rest mass, $\delta M\sim \alpha\hbar\omega_{0}/c^{2}\ll M$. In contrast, the term
$-\hbar\mathbf{k}\cdot\mathbf{p}/m_{e}$ generates a Doppler shift on the
photon frequency which depends on the relative direction between $\mathbf{k}$ and the current of the chomophoric electron, $\mathbf{j}_{e}=-e\mathbf{p}/m_{e}$. It is the break down of
time reversal and mirror symmetry that makes $\mathbf{p}$ take a non-vanishing transient value \footnote{By transient value of the operator $\mathbf{p}$
we mean the quantum amplitude of $\mathbf{p}$ evaluated between the ground state and an intermediate state, say $|I\rangle$,
$\langle I|\mathbf{p}|\tilde{\Omega}_{0}\rangle$. The quantum amplitude $\langle\mathbf{p}\rangle$ in the expression for $\Delta\omega$ must be intended this way.}
in the direction parallel to the external magnetic field.  It is this
term which generates the spectral non-reciprocity for photons propagating in opposite directions,
$\Delta\omega(\mathbf{k})=\tilde{\omega}(\mathbf{k})-\tilde{\omega}(-\mathbf{k})\sim2\mathbf{k}\cdot\langle\mathbf{p}\rangle/m_{e}$, needed for the transfer of (transverse) linear momentum
from the vacuum field to matter.\\
\indent The denominator of Eq.(\ref{Pperpo}) must be expanded up to first order in the Doppler shift term. At this order $\hbar\mathbf{k}\cdot\mathbf{p}/m_{e}$
is the non-relativistic Doppler shift, which does not clash with our non-relativistic approach \cite{RecoilRelativ}. This yields an even factor of $\mathbf{k}$ in
the integrand, necessary for the integral not to vanish. The term $\frac{\hbar\mu}{\mu^{*}m_{e}}(\mathbf{r}\wedge\mathbf{k})\cdot\mathbf{B}_{0}$
generates a vanishing contribution. Disregarding the mass renormalization terms, we find,
\begin{align}
\langle\mathbf{P}^{\textrm{Cas}}_{\perp}\rangle&=\Re{}\frac{\hbar^{3} e^{2}}{cm_{e}^{3}\epsilon_{0}}
\int\frac{\textrm{d}^{3}k\:\mathbf{k}}{(2\pi)^{3}k}\langle\tilde{\Omega}_{0}|
(\mathbf{p}+\frac{e}{2}\mathbf{B}_{0}\wedge\mathbf{r})\nonumber\\&\times\frac{\cdot(\mathbb{I}-\frac{\mathbf{k}\otimes\mathbf{k}}{k^{2}})\cdot}
{(\hbar^{2}k^{2}/2m_{e}+\hbar ck-E_{0}+\tilde{H}_{0})^{2}}(\mathbf{k}\cdot\mathbf{p})\nonumber\\
&\times\frac{1}{\hbar^{2}k^{2}/2m_{e}+\hbar ck-E_{0}+\tilde{H}_{0}}(\mathbf{p}+\frac{e}{2}\mathbf{B}_{0}\wedge\mathbf{r})|\tilde{\Omega}_{0}\rangle.\label{Pperpotro}
\end{align}
At the same time, since both the chiral interaction $V_{C}$ and the Zeeman potential $V_{Z}$ featuring in $\tilde{H}_{0}$ are treated as perturbations to
the harmonic oscillator, the denominator is to be expanded up to order $V_{C}V_{Z}$ and only terms at $\mathcal{O}(C$B$_{0})$ must be retained. The calculation
includes a number of finite integrals of the form,
\begin{align}
\Re&\int_{0}^{\infty}k^{3}\textrm{d}k[\frac{1}{(E^{m_{e}}_{k}-E_{1})(E^{m_{e}}_{k}-E_{2})^{2}}\nonumber\\&+
\frac{1}{(E^{m_{e}}_{k}-E_{2})(E^{m_{e}}_{k}-E_{1})^{2}}]\nonumber\\
&\simeq\frac{2m^{2}_{e}}{\hbar^{4}(E_{1}-E_{2})}\Bigl[\log{(E_{1}/E_{2})}+\mathcal{O}(\hbar\omega_{0}/m_{e}c^{2})\nonumber\\&+\mathcal{O}[(\hbar\omega_{0}/m_{e}c^{2})^{2}]+...\Bigr],
\label{integral}
\end{align}
where $E^{m_{e}}_{k}=\hbar^{2}k^{2}/2m_{e}+\hbar ck$ and $E_{1,2}$ are the energies of the electronic transitions between intermediate states, which are of the order of
$\hbar\omega_{0}$. Additional integrals involving the product of three and four fractions appear also in Eq.(\ref{Pperpotro}) which yield terms of the same orders.
The final expression can be greatly simplified assuming small anisotropy factors, $\eta_{ij}\ll1$. Averaging over the molecule's orientations ($rot$)
\cite{Craigbook} and expanding the result up to leading order in the anisotropy factors we obtain,
\begin{align}
\langle\mathbf{P}^{\textrm{Cas}}_{\perp}\rangle_{rot}&=[20736\ln{(4/3)}-12928\ln{(2)}-14511]\nonumber\\
&\times\frac{Ce^{3}\mathbf{B}_{0}}{93312\pi^{2}c\epsilon_{0}m_{e}^{2}\omega_{x}\omega_{y}\omega_{z}}
\eta^{zy}\eta^{xz}\eta^{yx}\nonumber\\&\simeq
\frac{-1.06Ce^{3}\mathbf{B}_{0}}{144\pi^{2}c\epsilon_{0}m_{e}^{2}\omega_{x}\omega_{y}\omega_{z}}
\eta^{zy}\eta^{xz}\eta^{yx}.\label{Pperptot}
\end{align}
\indent It is clear that $\langle\mathbf{P}^{\textrm{Cas}}_{\perp}\rangle_{rot}$ is the expression to be compared with the semiclassical result of Eq.(\ref{PCascalss}),
since from Eq.(\ref{ec2}) we read that the semiclassical $\mathbf{P}^{\textrm{Cas}}$  is the momentum of the radiative (i.e., transverse) modes only. First
we observe that $\langle\mathbf{P}^{\textrm{Cas}}_{\perp}\rangle_{rot}$ is a factor $(m_{e}c^{2}/\hbar\omega_{0})^{2}$ times larger than the semiclassical
$\mathbf{P}^{\textrm{Cas}}$. From Eq.(\ref{integral}) we have that the semiclassical result is included in our quantum calculation
(there are no other terms of the same order in the diagrams neglected), but it enters at a second order
correction, $\mathcal{O}[(m_{e}c^{2}/\hbar\omega_{0})^{2}]$, to the leading term. We conclude that the reason for this discrepancy is the failure of the semiclassical approach to account for the Doppler effect due to the relative motion of the internal charges in order to generate the necessary spectral non-reciprocity. Instead, spectral non-reciprocity comes in the semiclassical approach from the effective magnetochiral birefringence. 
We also note that the result of Eq.(\ref{Pperptot}) assumes an integration over $k$ up to infinity in Eq.(\ref{Pperpotro}), which violates the non-relativistic limit. However, had we assumed a cut-off for
$k_{max}$ of the order of the inverse of the Compton electronic wavelength, $k_{max}\sim m_{e}c/\hbar$, the expression on the r.h.s. of Eq.(\ref{Pperptot})
would change by only a factor of order unity.

\subsection{Longitudinal Casimir momentum}
For the computation of $\langle\mathbf{P}_{\parallel}^{\textrm{Cas}}\rangle$ the details of the calculation have been already published
in Refs.\cite{PRLDonaire} and the Supplemental Material there. Here we concentrate on the underlying mechanism which originates the transfer of momentum from the vacuum
field to the molecule. The calculation at $\mathcal{O}(C$B$_{0})$ implies up to third-order perturbation theory
in $V_{Z}+V_{C}+W$ to the ground state of the harmonic oscillator Hamiltonian, $|0\rangle$,
\begin{equation}
\langle\mathbf{P}^{\textrm{Cas}}_{\parallel}\rangle=\sum_{\mathbf{Q},I,\gamma_{\mathbf{k}\mathbf{\epsilon}}}
\frac{\langle\mathbf{Q}_{0},\tilde{\Omega}_{0}|e\Delta\mathbf{A}|\mathbf{Q},I,\gamma\rangle
\langle\mathbf{Q},I,\gamma|\tilde{W}|\mathbf{Q}_{0},\tilde{\Omega}_{0}\rangle}{\mathbf{Q}^{2}_{0}/2M+E_{0}-E_{Q,I,k}}+c.c.,\nonumber
\end{equation}
where $|\mathbf{Q}_{0},\tilde{\Omega}_{0}\rangle=\exp{(i\mathbf{Q}_{0}\cdot\mathbf{R}/\hbar)}|\tilde{\Omega}_{0}\rangle$ and
$\Delta\mathbf{A}=\mathbf{A}(\mathbf{r}_{N})-\mathbf{A}(\mathbf{r}_{e})$.
The $\mathbf{Q}_{0}$ dependent terms are shown in Ref.\cite{PRLDonaire} to give rise to mass renormalization factors. As for the rest we have,
\begin{align}\label{Plong}
\langle\mathbf{P}^{\textrm{Cas}}_{\parallel}\rangle&=\Bigl[\frac{-\hbar e}{2c\epsilon_{0}}\int\frac{\textrm{d}^{3}k}{(2\pi)^{3}k}\langle\tilde{\Omega}_{0}|
\frac{(\mathbb{I}-\frac{\mathbf{k}\otimes\mathbf{k}}{k^{2}})\cdot}{\hbar^{2}k^{2}/2m_{e}+\hbar ck-E_{0}+\tilde{H}_{0}}\nonumber\\
&e\mathbf{p}/m_{e}|\tilde{\Omega}_{0}\rangle+c.c.\Bigr]-[m_{e}\rightarrow m_{N}],
\end{align}
where $-[m_{e}\leftrightarrow m_{N}]$ means that the same expression within the square brackets must be evaluated exchanging
$m_{e}$ with $m_{N}$ and subtracted. In contrast to $\langle\mathbf{P}^{\textrm{Cas}}_{\perp}\rangle$, no Doppler shift term enters the calculation and no spectral
non-reciprocity exists. Nonetheless the longitudinal momentum is also generated by the non-vanishing transient currents parallel to the external magnetic field.
In this case, however, the currents are due to the internal motion of both the chromophoric electron and the nucleus, $\mathbf{j}_{e,N}=e\mathbf{p}/m_{e,N}$.
They are the sources of the magnetic fields at some point $\mathbf{X}$ measured w.r.t. the center of mass,
\begin{equation}
\mathbf{B}_{e,N}(\mathbf{X})=\frac{-\mu_{0}}{4\pi}\mathbf{j}_{e,N}\wedge\frac{\mathbf{X}}{X^{3}}=
-\frac{i\mu_{0}e}{m_{e,N}}\int\frac{\textrm{d}^{3}}{(2\pi)^3}e^{i\mathbf{k}\cdot\mathbf{X}}\frac{\mathbf{k}\wedge\mathbf{p}}{k^{2}}.\label{Bfield}
\end{equation}
Each of these magnetic fields, combined with the Coulomb field ($Coul$) of the respective charges, $+e$ for the nucleus and $-e$ for the electron,
\begin{equation}
\mathbf{E}^{Coul}_{e,N}(\mathbf{X})=\frac{\pm e}{4\pi\epsilon_{0}}\frac{\mathbf{X}}{X^{3}}=
\mp\frac{ie}{\epsilon_{0}}\int\frac{\textrm{d}^{3}k}{(2\pi)^3}e^{i\mathbf{k}\cdot\mathbf{X}}\frac{\mathbf{k}}{k^{2}},
\end{equation}
give rise to the longitudinal Casimir momentum of Eq.(\ref{Plong}),
\begin{align}
\langle\mathbf{P}^{\textrm{Cas}}_{\parallel}\rangle&=\epsilon_{0}\int\textrm{d}^{3}X\langle\mathbf{E}^{Coul}_{e}(\mathbf{X})\wedge\mathbf{B}^{*}_{e}(\mathbf{X})\rangle
\nonumber\\&+\epsilon_{0}\int\textrm{d}^{3}X\langle\mathbf{E}^{Coul}_{N}(\mathbf{X})\wedge\mathbf{B}^{*}_{N}(\mathbf{X})\rangle.\label{Plongy}
\end{align}
In order to obtain an equivalence between equations (\ref{Plong}) and (\ref{Plongy}), one of the $k$ factors in the Fourier transform of the magnetic fields in
Eq.(\ref{Bfield}) must be
substituted by $k+\hbar k^{2}/2cm_{e,N}-(E_{0}-\tilde{H}_{0})/\hbar c$, where the momentum shift comes from the sum over the intermediate states
which enter the transient currents. It is clear that the only difference between the two terms of Eqs.(\ref{Plong},\ref{Plongy}) is in the sign of the charges which
enter the electrostatic fields and the masses which determine their respective currents. As a result, the addition of both terms must be proportional
to $f(m_{e})-f(m_{N})$, with $f$ a real functional of the masses. The result is,
\begin{equation}\label{Pkinparx}
\langle \mathbf{P}^{\textrm{Cas}}_{\parallel}\rangle=\frac{Ce^{3}\ln{(m_{N}/m_{e})}}{96\pi^{2}c\epsilon_{0}\mu\mu^{*}
(\omega_{x}+\omega_{y}+\omega_{z})}\sum_{i,j,k}\varepsilon_{ijk}\frac{\textrm{B}_{0}^{i}\eta^{kj}}{\omega_{k}\omega_{j}}\hat{i},
\end{equation}
where $\varepsilon_{ijk}$ is the three-dimensional Levi-Civita tensor, the indices $i,j,k$ take on the axis labels $x,y,z$ and $\hat{i}$ is a unitary vector along the $i$ axis. After averaging over the molecule orientations
we end up with,
\begin{equation}\label{PkinparAverage}
\langle\mathbf{P}^{\textrm{Cas}}_{\parallel}\rangle_{rot}=\frac{Ce^{3}\ln{(m_{e}/m_{N})}\mathbf{B}_{0}}{144\pi^{2}c\epsilon_{0}\mu\mu^{*}\omega_{x}\omega_{y}\omega_{z}}
\eta^{zy}\eta^{xz}\eta^{yx}.
\end{equation}
As for the transverse Casimir momentum, the above calculation holds if we take the upper limit of integration of Eq.(\ref{Plong}) at infinity.
Again, for a cut-off of the order of $k_{max}\sim m_{e}c/\hbar$, additional terms of the same order as those in
Eq.(\ref{Pkinparx}) would have been obtained.

\subsection{Casimir momentum as a function of optical parameters}

Finally, adding up the transverse and the longitudinal contributions we obtain,
\begin{equation}\label{PkinparAverage}
\langle\mathbf{P}^{\textrm{Cas}}\rangle_{rot}=\frac{Ce^{3}\mathbf{B}_{0}[\ln{(m_{e}/m_{N})}+1]}{144\pi^{2}c\epsilon_{0}\mu\mu^{*}\omega_{x}\omega_{y}\omega_{z}}
\eta^{zy}\eta^{xz}\eta^{yx}.
\end{equation}
This formula is a simple expression for $\langle\mathbf{P}^{\textrm{Cas}}\rangle_{rot}$
in terms of the chiral parameter $C$, the magnetic field and the natural frequencies of the oscillator. We can write it also
in terms of the fine structure constant, $\alpha$, the static optical rotatory power,  $\beta(0)$, and the static electric polarizability of the molecule,
$\alpha_{E}(0)$.
 By comparing the above expression with the formulas for $\beta$ and $\alpha_{E}$ in \ref{appendB} we obtain,\footnote{We have fixed here an erroneous
 minus sign appearing in Ref.\cite{PRLDonaire}.}
\begin{equation}\label{main1}
\langle\mathbf{P}^{\textrm{Cas}}\rangle_{rot}\simeq\frac{2\alpha}{9\pi}\frac{\beta(0)}{\alpha_{E}(0)}[\ln{(m_{N}/m_{e})}+1]e\mathbf{B}_{0}.
\end{equation}
In Ref.\cite{PRLDonaire} we speculate that, apart from constants of order unity, this expression is model-independent. For a given set of natural frequencies, all the atomic lengths in the
problem are determined by quantum mechanics. In particular, $\beta(0)/ \alpha_{E}(0)$ is a length, a fraction of the electronic Compton wavelength,
that we identify with a chiral length, $l_{ch}$. Therefore, we can write $\textrm{P}^{\textrm{Cas}}\sim\alpha\:e$B$_{0}l_{ch}$ and thus interpret it as the leading QED correction to the classical Abraham momentum.

\section{Origin of  Casimir-induced kinetic energy}\label{sec5}

\indent In the precedent sections we have found that the chiral
molecule acquires a kinetic momentum during the switching of the
external magnetic field as a result of its interaction with the
vacuum field. On the other hand we have shown that, as a result of the
conservation of the total momentum $\mathbf{K}$, there exists a transfer of
linear momentum from the vacuum field to the molecule. The question arises
 whether the resultant variation of kinetic energy is provided by the vacuum field or
by an external source. In the following we prove that this
variation of kinetic energy is part of the energy provided or removed by the source of
the external magnetic field to generate a vacuum correction to the molecular magnetization. That magnetization energy can be considered as part of the 'magnetic Lamb energy'.\\
\indent To this aim we make use of the Hellmann-Feynman-Pauli
theorem \cite{Noztier}. According to this theorem, the variation of the total
energy of a system in its ground state with respect to an adiabatic parameter,
$\lambda$, can be computed from the expectation value of the
functional derivative of the Hamiltonian w.r.t. that parameter. The
variation of the total energy is the work done over the system of
interest by an external source responsible for the adiabatic
variation of the parameter. If the parameter enters an interaction
potential, $W_{int}$, that work reads
\begin{equation}
\mathcal{W}_{\lambda}=\int_{0}^{\lambda}\delta\lambda' \langle \delta W_{int}/\delta\lambda'\rangle.\label{We}
\end{equation}
For instance, this approach is used to compute intermolecular forces by
considering their center of mass position vectors as adiabatic
parameters \cite{Zhang}.\\
\indent In our case, the adiabatic parameter to be varied is the magnetic field,
$\mathbf{B}_{0}$, and the variation of energy to be computed is a magnetic energy, $\mathcal{W}_{\mathbf{B}_{0}}$.
Since $\mathbf{B}_{0}$ enters  $V_{Z}$ and $W$ linearly, and $\Delta V$ quadratically in Eq.(\ref{effective}) \footnote{The term of order
$QB_{0}$ in $\Delta V$, $e\mathbf{Q}\cdot(\mathbf{r}\wedge\mathbf{B}_{0})/M$, does not contribute to $\mathcal{W}_{\mathbf{B}_{0}}$.}, the variation of the total energy
of the system molecule-EM vacuum during the adiabatic switching of the
magnetic field from $\mathbf{0}$ up to its final value
$\mathbf{B}_{0}$ is, 
\begin{align}
\mathcal{W}_{\mathbf{B}_{0}}&=\int_{\mathbf{0}}^{\mathbf{B}_{0}}\delta\mathfrak{B}_{0}
\langle\delta(\tilde{W}+V_{Z}+\Delta V)/\delta\mathfrak{B}_{0}\rangle\nonumber\\&=\int_{\mathbf{0}}^{\mathbf{B}_{0}}\delta\mathfrak{B}_{0}
[\langle \tilde{W}_{\mathfrak{B}_{0}}+V_{Z}\rangle/\mathfrak{B}_{0}+2\langle \Delta V\rangle/\mathfrak{B}_{0}],\label{Work}
\end{align}
where in the last equality  $\tilde{W}_{\mathfrak{B}_{0}}=(e^{2}/2)(\mathbf{r}\wedge\mathfrak{B}_{0})\cdot[\mathbf{A}(\mathbf{R}-\frac{m_{N}}{M}\mathbf{r})/m_{e}-
\mathbf{A}(\mathbf{R}+\frac{m_{e}}{M}\mathbf{r})/m_{N}]$
includes only the $\mathfrak{B}_{0}$-dependent terms of $\tilde{W}$. We note that $\mathcal{W}_{\mathbf{B}_{0}}$ is the work done by the external source that
generates the uniform magnetic field $\mathbf{B}_{0}$ in the space occupied by the chiral molecule in the presence of the quantum vacuum. The expression
within the square brackets in Eq.(\ref{Work}) is indeed the total magnetization of the molecule in its ground state, $\langle \mathbf{M}_{\mathfrak{B}_{0}}\rangle$,
with a minus sign in front. Hence we can write,
$\mathcal{W}_{\mathbf{B}_{0}}=-\int_{\mathbf{0}}^{\mathbf{B}_{0}}\langle \mathbf{M}_{\mathfrak{B}_{0}}\rangle\cdot\delta\mathfrak{B}_{0}$.\\
\indent Up to our level of approximation
$\mathcal{W}_{\mathbf{B}_{0}}$ must include terms of up to order two in $C$, $\mathbf{B}_{0}$ and $\alpha$. It follows that perturbation theory has to be applied
up to order six in $V_{C}+V_{Z}+\Delta V+\tilde{W}$. We first observe that at leading order in $\Delta V$ and quadratic in $V_{Z}$, and at zero order in $V_{C}$ and
$\tilde{W}$, $\mathcal{W}_{\mathbf{B}_{0}}$ reduces to the magnetization energy
 of the diamagnetic molecule in the absence of vacuum field,
$\sim-\alpha_{M}(0)B_{0}^{2}$, $\alpha_{M}(0)$ being the static
magnetic polarizability given in \ref{appendB}. At higher order in perturbation theory, including the chiral potential and the vacuum field,
we find a number of additional terms in $\mathcal{W}_{\mathbf{B}_{0}}$. Most of these terms
amount to variations in the energy of internal atomic levels, which are not relevant to us. We are however
interested in those terms of $\mathcal{W}_{\mathbf{B}_{0}}$ which
are of the order of the variation of the kinetic energy of the chiral group,
$\Delta E_{\textrm{kin}}(\mathbf{B}_{0})=[\mathbf{Q}_{0}-\langle\mathbf{P}^{\textrm{Cas}}(\mathbf{B}_{0})\rangle]^{2}/2M-\mathbf{Q}^{2}_{0}/2M$, for an initial kinetic
momentum $\mathbf{Q}_{0}$ at zero magnetic field.
In the following we search for these terms by inspection and
show that altogether they add up to give $\Delta E_{\textrm{kin}}$. As a result, we can say that the variation of the kinetic energy is supplied by the source of the external magnetic field to produce a vacuum correction in the magnetization of the chiral molecule.\\
\indent First we note that keeping finite the momentum of the molecule, $\mathfrak{Q}_{0}$,
and bearing in mind that during the adiabatic switching of the magnetic field $\mathfrak{Q}_{0}$ equals $\mathbf{Q}_{0}-\langle\mathbf{P}^{\textrm{Cas}}(\mathfrak{B}_{0})\rangle$,
$\mathcal{W}_{\mathbf{B}_{0}}$ contains terms quadratic in $C$, $\mathbf{B}_{0}$ and $\alpha$ from the application of perturbation theory
to $e^{-i\mathfrak{Q}_{0}\cdot\mathbf{R}/\hbar}|0\rangle$ up to order one in $V_{C}$ and $V_{Z}$,
and up to order two in $\tilde{W}$, as for the computation of the Casimir momentum. Following the same steps as for the calculation of $\langle\mathbf{P}^{\textrm{Cas}}\rangle$,
we start with the dressed ground state, $|\mathfrak{Q}_{0},\tilde{\Omega}_{0}\rangle=e^{-i\mathfrak{Q}_{0}\cdot\mathbf{R}/\hbar}|\tilde{\Omega}_{0}\rangle$, and apply to it second order perturbation theory with the interaction
potential $\tilde{W}$. The calculation is therefore similar to that of the
ordinary Lamb shift except for the fact that the ground state and the intermediate states in our case are dressed up to first order in $C$ and $\mathfrak{B}_{0}$. In addition
to the diagram of Fig.\ref{fig1}i($b$) for the self-energy of the electron we
 must add the one of Fig.\ref{fig1}i($a$) for the nucleus. We will denote the resultant energy by $E^{Lamb}(\mathfrak{B}_{0})$ and refer to it as magnetic Lamb energy,
\begin{equation}
E^{Lamb}(\mathfrak{B}_{0})=\sum_{\mathbf{Q},I,\gamma_{\mathbf{k}\mathbf{\epsilon}}}
\frac{\langle\mathfrak{Q}_{0},\tilde{\Omega}_{0}|\tilde{W}|\mathbf{Q},I,\gamma\rangle
\langle\mathbf{Q},I,\gamma|\tilde{W}|\mathfrak{Q}_{0},\tilde{\Omega}_{0}\rangle}{\mathfrak{Q}^{2}_{0}/2M+E_{0}-E_{Q,I,k}}.
\end{equation}
Let us reorganize first the terms in $\tilde{W}$ as
\begin{align}\label{tW}
\tilde{W}&=-\frac{e}{m_{N}}\left(\mathbf{p}-\frac{e}{2}\mathfrak{B}_{0}\wedge\mathbf{r}\right)\cdot
\mathbf{A}(\mathbf{r}_{N})\\
&-\frac{e}{m_{e}}\left(\mathbf{p}+\frac{e}{2}\mathfrak{B}_{0}\wedge\mathbf{r}\right)\cdot
\mathbf{A}(\mathbf{r}_{e})-\frac{\mathbf{P}\cdot\mathbf{P}_{\parallel}^{\textrm{Cas}}}{M}+\mathcal{O}(\textrm{A}^{2}).\nonumber
\end{align}
This shows that the longitudinal Casimir momentum appears naturally
coupled to the canonical momentum of the center of mass. In the following we disregard the terms of $E^{Lamb}$ associated to internal energies and
restrict ourselves to those of the order of $E_{\textrm{kin}}$. In the first place, from the combination of the factors $-\mathbf{p}\cdot[\frac{e}{m_{N}}\mathbf{A}(\mathbf{r}_{N})+\frac{e}{m_{e}}\mathbf{A}(\mathbf{r}_{e})]$
and $-M^{-1}\mathbf{P}\cdot\mathbf{P}_{\parallel}^{\textrm{Cas}}$ in $\tilde{W}^{2}$ we have,
\begin{align}
 E_{\parallel}^{Lamb}&=\big[-\sum_{\mathbf{Q},I,\gamma_{\mathbf{k}\mathbf{\epsilon}}}
\frac{\langle\mathfrak{Q}_{0},\tilde{\Omega}_{0}|e\mathbf{p}\cdot\mathbf{A}(\mathbf{r}_{e})/m_{e}|\mathbf{Q},I,\gamma\rangle}{\mathfrak{Q}^{2}_{0}/2M+E_{0}-E_{Q,I,k}}\nonumber\\
&\times\langle\mathbf{Q},I,\gamma|e\mathbf{A}(\mathbf{r}_{e})\cdot\mathbf{P}/M|\mathfrak{Q}_{0},\tilde{\Omega}_{0}\rangle+c.c.\big]\nonumber\\
&-[m_{e}\rightarrow m_{N}].
\end{align}
Since the ground state $|\mathfrak{Q}_{0},\tilde{\Omega}_{0}\rangle$ is an eigenstate of the center of mass canonical momentum,
$\mathbf{P}|\mathfrak{Q}_{0},\tilde{\Omega}_{0}\rangle=\mathfrak{Q}_{0}|\mathfrak{Q}_{0},\tilde{\Omega}_{0}\rangle$, the above equation can be written as
\begin{align}
 E_{\parallel}^{Lamb}&=\big[-\sum_{\mathbf{Q},I,\gamma_{\mathbf{k}\mathbf{\epsilon}}}
\frac{\langle\mathfrak{Q}_{0},\tilde{\Omega}_{0}|e\mathbf{p}\cdot\mathbf{A}(\mathbf{r}_{e})/m_{e}|\mathbf{Q},I,\gamma\rangle}
{\mathfrak{Q}^{2}_{0}/2M+E_{0}-E_{Q,I,k}}\nonumber\\
&\times\langle\mathbf{Q},I,\gamma|e\mathbf{A}(\mathbf{r}_{e})|\mathfrak{Q}_{0},\tilde{\Omega}_{0}\rangle\cdot\mathfrak{Q}_{0}/M+c.c.\big]\nonumber\\
&-[m_{e}\rightarrow m_{N}].
\end{align}
The expression under the sum is exactly the one for the longitudinal Casimir momentum in Eq.(\ref{Plong}) with a minus sign in front. Therefore we find,
\begin{equation}
 E_{\parallel}^{Lamb}=-\langle\mathbf{P}^{\textrm{Cas}}_{\parallel}(\mathfrak{B}_{0})\rangle\cdot\mathfrak{Q}_{0}/M.\label{Eparal}
\end{equation}
This relation explains the subscript $\parallel$ in $ E_{\parallel}^{Lamb}$.\\
\indent Likewise, from the combination of two factors $-\frac{e}{m_{e}}\left(\mathbf{p}+\frac{e}{2}\mathfrak{B}_{0}\wedge\mathbf{r}\right)\cdot
\mathbf{A}(\mathbf{r}_{e})$ in $\tilde{W}^{2}$ we have,
\begin{align}
&\sum_{\mathbf{Q},I,\gamma_{\mathbf{k}\mathbf{\epsilon}}}
\frac{|\langle\mathfrak{Q}_{0},\tilde{\Omega}_{0}|\frac{e}{m_{e}}\left(\mathbf{p}+\frac{e}{2}\mathfrak{B}_{0}\wedge\mathbf{r}\right)\cdot
\mathbf{A}(\mathbf{r}_{e})|\mathbf{Q},I,\gamma\rangle|^{2}}
{\mathfrak{Q}^{2}_{0}/2M+E_{0}-E_{Q,I,k}}
\nonumber\\&=\frac{-\hbar e^{2}}{2cm_{e}^{2}\epsilon_{0}}
\int\frac{\textrm{d}^{3}k}{(2\pi)^{3}k}\langle\tilde{\Omega}_{0}|
(\mathbf{p}+\frac{e}{2}\mathfrak{B}_{0}\wedge\mathbf{r})\nonumber\\
&\times\frac{\cdot(\mathbb{I}-\frac{\mathbf{k}\otimes\mathbf{k}}{k^{2}})\cdot}{\hbar^{2}k^{2}/2m_{e}+\hbar ck-\hbar\mathbf{k}\cdot\mathbf{p}/m_{e}
-\hbar\mathbf{k}\cdot\mathfrak{Q}_{0}/M-E_{0}+\tilde{H}_{0}}\nonumber\\
&\times(\mathbf{p}+\frac{e}{2}\mathfrak{B}_{0}\wedge\mathbf{r})|\tilde{\Omega}_{0}\rangle.\label{Pperpal}
\end{align}
Next we note in the denominator a Doppler shift term due to the
motion of the center of mass,
$-\hbar\mathbf{k}\cdot\mathfrak{Q}_{0}/M$. Expanding the
fraction up to first order in this term and discarding the
zero-order terms we find,
\begin{align}
 E^{\perp}_{Lamb}&=\frac{-\hbar e^{2}}{2cm_{e}^{2}\epsilon_{0}}
\int\frac{\textrm{d}^{3}k}{(2\pi)^{3}k}\langle\tilde{\Omega}_{0}|\frac{\hbar}{M}(\mathbf{k}\cdot\mathfrak{Q}_{0})
(\mathbf{p}+\frac{e}{2}\mathfrak{B}_{0}\wedge\mathbf{r})\nonumber\\
&\times\frac{\cdot(\mathbb{I}-\frac{\mathbf{k}\otimes\mathbf{k}}{k^{2}})\cdot}{(\hbar^{2}k^{2}/2m_{e}+\hbar ck-\hbar\mathbf{k}\cdot\mathbf{p}/m_{e}
-E_{0}+\tilde{H}_{0})^{2}}\nonumber\\
&\times(\mathbf{p}+\frac{e}{2}\mathfrak{B}_{0}\wedge\mathbf{r})|\tilde{\Omega}_{0}\rangle.\label{Pperpal}
\end{align}
Lastly, by expanding the denominator up to first order in $-\hbar\mathbf{k}\cdot\mathbf{p}/m_{e}$ and comparing with Eq.(\ref{Pperpotro}) we end up with
\begin{equation}
 E^{\perp}_{Lamb}=-\langle\mathbf{P}^{\textrm{Cas}}_{\perp}(\mathfrak{B}_{0})\rangle\cdot\mathfrak{Q}_{0}/M.\label{Eperp}
\end{equation}
\indent We have found in Sec.\ref{Qapproach} that
$\langle\mathbf{P}^{\textrm{Cas}}(\mathfrak{B}_{0})\rangle\propto\mathfrak{B}_{0}$
in stationary conditions. Therefore, combining the results of
equations (\ref{Eparal}) and (\ref{Eperp}) and considering  Eq.(\ref{Work}) we can write the magnetic Lamb energy associated to these two terms as,
\begin{equation}\label{ener}
\int_{0}^{\mathbf{B}_{0}}\delta\mathfrak{B}_{0}(E_{\parallel}^{Lamb}+
E_{\perp}^{Lamb})/\mathfrak{B}_{0}=
\langle\mathbf{P}^{\textrm{Cas}}(\mathbf{B}_{0})\rangle^{2}/2M-\mathbf{Q}_{0}\cdot\langle\mathbf{P}^{\textrm{Cas}}(\mathbf{B}_{0})\rangle/M,
\end{equation}
which equals $\Delta E_{\textrm{kin}}(\mathbf{B}_{0})=[\mathbf{Q}_{0}-\langle\mathbf{P}^{\textrm{Cas}}(\mathbf{B}_{0})\rangle]^{2}/2M-\mathbf{Q}^{2}_{0}/2M$
as anticipated. From this result we see that the variation of the kinetic energy of the chiral group, induced by
the transfer of linear momentum from the vacuum to the molecule, has its origin in the magnetic Lamb energy
(i.e., that part of the interaction energy that is  induced  by both the magnetic field and the quantum vacuum)\footnote{In our simplified model the chiral group is free to move within the molecule. The subsequent transfer of kinetic momentum from the chiral group
to the rest of the molecule would be accompanied by a redistribution of mechanical energy.}. This kinetic energy is
the magnetic energy associated with the vacuum correction to the magnetization of the molecule,
$\Delta M_{\mathfrak{B}_{0}}=-(E_{\parallel}^{Lamb}+
E_{\perp}^{Lamb})/\mathfrak{B}_{0}$, $\Delta E_{\textrm{kin}}(\mathbf{B}_{0})=-\int_{0}^{\mathbf{B}_{0}}\Delta M_{\mathfrak{B}_{0}}\delta\mathfrak{B}_{0}$ \footnote{$\Delta M_{\mathfrak{B}_{0}}=-(E_{\parallel}^{Lamb}+
E_{\perp}^{Lamb})/\mathfrak{B}_{0}$ is a shortcut to write the vacuum correction to the magnetization, and so is the expression in the integrand of Eq.(\ref{Work}) for writing the total magnetization. Strictly speaking it should be written $\Delta M_{\mathfrak{B}_{0}}=-\delta(E_{\parallel}^{Lamb}+E_{\perp}^{Lamb})/\delta\mathfrak{B}_{0}+(\mathfrak{B}_{0}/2)\delta^{2}(E_{\parallel}^{Lamb}+E_{\perp}^{Lamb})/\delta\mathfrak{B}_{0}^{2}$.}. The sign of this energy depends on the magnitudes
of the initial momentum and final magnetic field, $Q_{0}$ and $B_{0}$, as well as on their relative orientation. A positive sign means that the energy
is provided by the external source which generates the magnetic field. A negative sign means that part of the initial kinetic energy of the molecule is
removed by the external source during the magnetization process.

\section{Conclusions}\label{sec6}

We have derived an expression for the Casimir momentum transferred from the EM vacuum to a chiral molecule during the switching of an external magnetic field.
We have modeled the chiral molecule using the single quantum oscillator model of Condon \emph{et al.} \cite{Condon1,Condon2,EPJDonaire}. We have applied both a
semiclassical approach and a fully quantum non-relativistic approach.\\
\indent The quantum approach reveals that two distinct mechanisms operate in the transfer of momentum, they both based on the production of transient internal
currents in the direction parallel to the external magnetic field. The transverse Casimir momentum [Eq.(\ref{Pperpotro})] has its origin in the spectral
non-reciprocity generated by a
Doppler shift in the frequency of the vacuum photons. This shift is due to the internal momentum of the chromophoric electron, $\mathbf{p}$. The break down of
time reversal and mirror symmetry makes $\mathbf{p}$ take a non-vanishing transient value
parallel to the external magnetic field. On the contrary, the longitudinal Casimir momentum [Eq.(\ref{Plong})] has its origin in the combination of the
electrostatic field of the internal charges and the magnetic field generated by their transient currents parallel to the external
magnetic field [Eq.(\ref{Plongy})]. Therefore, we have found that the longitudinal Casimir momentum is the momentum of the source EM field while the transverse momentum is the momentum of the sourceless vacuum field. In particular, the finding  $\langle\mathbf{P}^{\textrm{Cas}}_{\perp}\rangle\neq\mathbf{0}$ proves that by modifying the symmetries of the space-time it is possible to vary the vacuum expectation value of observable quantities.\\
\indent The quantum result in Eq.(\ref{main1}) is linear in $\mathbf{B}_{0}$ and proportional to the fine structure constant, $\alpha$, and to
the molecular rotatory power.  It is conjectured that this result is universal and model-independent, up to numerical prefactors of order
unity which might include relativistic corrections.\\
\indent The semiclassical approach only applies to the calculation of the transverse Casimir momentum and relies on the non-reciprocity of the spectrum of
effective normal modes. However, it fails to incorporate the net effect of the Doppler shift which comes in into the microscopical quantum approach and which is seen to dominate largely. The semiclassical result of Eq.(\ref{PCascalss}) appears as a second order correction to the leading order term of the quantum result.\\
\indent We have proved that the variation of the kinetic energy of the chromophoric group has its origin in the magnetic Lamb energy [Eq.(\ref{ener})]. This kinetic energy is found to be part of
the magnetic energy associated with the vacuum correction to the magnetization of the molecule.
When positive, this energy is provided by the external source which generates the uniform magnetic field. When negative,
it is part of the initial kinetic energy of the molecule which is
removed by the external source during the magnetization process.\\

\ack

This work was supported by the ANR contract PHOTONIMPULS
ANR-09-BLAN-0088-01.

\appendix
\section{Rotationally averaged polarizabilities}\label{appendB}
\indent At leading order in the anisotropy factors, the polarizabilities that enter Eq.(\ref{dind}) are \cite{EPJDonaire},
\begin{align}
&\alpha_{E}=\frac{e^{2}}{\mu(\omega_{0}^{2}-\omega^{2})},\quad\alpha_{M}=\frac{4e^{2}\hbar\omega_{0}\mathcal{N}_{xyz}}{9\mu^{*2}(4\omega_{0}^{2}-\omega^{2})},\nonumber\\
&\chi=\frac{e^{3}}{\mu\mu^{*}(\omega_{0}^{2}-\omega^{2})^{2}},\quad\zeta=\frac{e^{3}\hbar\omega_{0}(4\omega_{0}^{2}-3\omega^{2})\mathcal{N}_{xyz}}
{18\mu^{*3}\omega(4\omega_{0}^{2}-\omega^{2})^{2}},\nonumber\\
&\beta=\frac{-2e^{2}\hbar C\omega^{3}_{0}(\omega^{4}+7\omega_{0}^{2}\omega^{2}+4\omega_{0}^{4})\mathcal{M}_{xyz}}
{\mu^{2}\mu^{*}(\omega^{4}-5\omega_{0}^{2}\omega^{2}+4\omega_{0}^{4})^{3}},\nonumber\\
&\gamma=\frac{e^{3}\hbar C\omega^{3}_{0}\omega^{2}(\omega^{2}+12\omega_{0}^{2})\mathcal{M}_{xyz}}
{\mu^{2}\mu^{*2}(\omega^{4}-5\omega_{0}^{2}\omega^{2}+4\omega_{0}^{4})^{3}},\nonumber\\
&\xi=\frac{-2e^{3}\hbar C\mathcal{M}_{xyz}\omega^{3}_{0}\omega(19\omega^{6}-842\omega^{4}\omega_{0}^{2}-224\omega^{2}\omega_{0}^{4}
-672\omega_{0}^{6})}
{15\mu^{2}\mu^{*2}(\omega^{2}-\omega_{0}^{2})^{3}(\omega^{2}-4\omega_{0}^{2})^{5}},\nonumber
\end{align}
where $\mathcal{M}_{xyz}$ and $\mathcal{N}_{xyz}$ are dimensionless functions of the anisotropy factors,
$\mathcal{M}_{xyz}\equiv\eta^{zy}\eta^{yx}\eta^{xz}$, $\mathcal{N}_{xyz}\equiv\eta^{yx}\eta^{yz}+\eta^{zx}\eta^{zy}+\eta^{xy}\eta^{xz}$.

\end{document}